\documentclass[namedreferences]{solarphysics}

\usepackage[hyperref,optionalrh]{spr-sola-addons} 
\usepackage{graphicx}        
\usepackage{amssymb}        
\usepackage{color}           
\usepackage[urlcolor=blue,breaklinks]{hyperref}
\hypersetup{pdfborder=0 0 0}

\ifx \doiurl    \undefined \def \doiurl#1{\href{http://dx.doi.org/#1}{\textsf{DOI}}}\fi
\ifx \adsurl    \undefined \def \adsurl#1{\href{http://adsabs.harvard.edu/abs/#1}{\textsf{ADS}}}\fi
\ifx \arxivurl  \undefined \def \arxivurl#1{\href{http://arxiv.org/abs/#1}{\textsf{arXiv}}}\fi

\newcommand{\etal}{{\it et al.}}

\begin{document}

\begin{article}

\begin{opening}

\title{Development of a Precise Polarization Modulator for UV Spectropolarimetry}

\author{S.~\surname{Ishikawa}$^{1,2}$\sep
        T.~\surname{Shimizu}$^{1}$\sep
        R.~\surname{Kano}$^{2}$\sep
        T.~\surname{Bando}$^{2}$\sep
        R.~\surname{Ishikawa}$^{2}$\sep
        G.~\surname{Giono}$^{2}$\sep
        S.~\surname{Tsuneta}$^{1}$\sep
        S.~\surname{Nakayama}$^{3}$\sep
        T.~\surname{Tajima}$^{3}$      
       }
\runningauthor{Ishikawa \etal}
\runningtitle{Development of a Precise Polarization Modulator for UV Spectropolarimetry}

   \institute{ $^{1}$ Institute of Space and Astronautical Science, Japan Aerospace Exploration Agency, 3-1-1 Yoshinodai, Chuo, Sagamihara, Kanagawa 252-5210, Japan\\
   email: \url{s.ishikawa@solar.isas.jaxa.jp}\\
    $^{2}$ National Astronomical Observatory of Japan, 2-21-1 Osawa, Mitaka, Tokyo 181-8588, Japan              
              \\
              $^{3}$ Mitsubishi Precision Co.,Ltd., Kamakura, Kanagawa, Japan
             }
\begin{abstract}
We developed a polarization modulation unit (PMU) to rotate a waveplate continuously 
in order to observe solar magnetic fields by spectropolarimetry.
The non-uniformity of the PMU rotation may cause errors in the measurement of the degree of linear polarization (scale error) and its angle (crosstalk between Stokes-$Q$ and -$U$), 
although it does not cause an artificial linear polarization signal (spurious polarization).  
We rotated a waveplate with the PMU to obtain a polarization modulation curve and estimated the scale error and crosstalk caused by the rotation non-uniformity.  
The estimated scale error and crosstalk were $<$0.01\,\% for both. 
This PMU will be used as a waveplate motor for the \textit{Chromospheric Lyman-Alpha SpectroPolarimeter} (CLASP) rocket experiment.
We confirmed that the PMU has the sufficient performance and function for CLASP.  
\end{abstract}
\end{opening}

\section{Introduction}
To understand the energy-release processes occurring in the Sun, it is important to measure the magnetic field, especially that in the upper layers of the solar atmosphere.  
The Zeeman effect is a powerful tool for revealing magnetic fields on the photosphere.  
However, it is difficult to apply the Zeeman effect for magnetic-field measurements in the upper chromosphere and above owing to weaker magnetic fields and larger thermal broadening in those layers.
The use of high-precision spectropolarimetry to detect the Hanle effect enables us to overcome this difficulty and infer the vector magnetic field 
\citep{hanle1924, trujillobueno2002, trujillobueno2011, belluzzi2012, ishikawa2014a}.

Spectropolarimetry to observe solar magnetic fields are performed both with 
ground-based telescopes -- such as 
the \textit{Tenerife Infrared Polarimeter} \citep{martinezpillet1999}, which is 
attached to the German \textit{Vacuum Tower Telescope} in Tenerife, and 
the \textit{Advanced Stokes Polarimeter} \citep{elmore1992} --
and with space telescopes such as the \textit{Solar Optical Telescope} \citep{tsuneta2008, lites2013} onboard the \textit{Hinode} spacecraft \citep{kosugi2007}.  
The Stokes parameters can be obtained by measuring the polarization modulations.
Although there are several ways to modulate the intensity, one of the widely used methods combines a continuously rotating waveplate and a polarization analyzer \citep{lites1987}.  
The continuously rotating waveplate offers advantages such as a simple control system and no dead time caused by the waveplate rotation.  
In this case, the non-uniformity of the continuous waveplate rotation causes errors in the polarization measurement.  

Errors in a polarization measurement can be classified into spurious polarization, scale error, and crosstalk among the Stokes parameters $Q$, $U$, and $V$ \citep{ichimoto2008}.  
The rotation non-uniformity of the waveplate rotator does not cause spurious polarization but may cause scale error or crosstalk.  
The scale error is the error in magnitude of each Stokes parameter: $Q$, $U$, and $V$.
It can be evaluated as $(p-p')/p$, where $p$ is the degree of polarization of the incident light, and $p'$ is its measured value.  

The SOLAR-C working group of the Institute of Space and Astronautical Science (ISAS), together 
with the Mitsubishi Precision Company, has been developing 
a motor to rotate a waveplate continuously for the \textit{Solar UV-Vis-IR Telescope} (SUVIT: \citealt{katsukawa2013}) onboard the next solar spacecraft SOLAR-C \citep{shimizu2014}.
We plan to use a development model of the motor for a rocket experiment named the \textit{Chromospheric Lyman-Alpha SpectroPolarimeter} (CLASP) to 
observe the linear polarization profile of the Lyman-$\alpha$ line (vacuum ultraviolet, 121.6~nm) from the Sun 
in order to perform magnetic-field measurements in the upper chromosphere and transition region \citep{kano2012, kobayashi2012}.  
In the CLASP instrument, the motor will continuously rotate a half-waveplate \citep{ishikawa2013}, and 
is required to have rotation uniformity for the linear polarization to be measured precisely.

The total tolerance of the scale error is 10\,\% for CLASP \citep{ishikawa2014b}.  
There are several sources of the scale error, and some of them contribute significantly (\textit{i.e.} the camera readout nonlinearity and scattered light contribute a few percent).  
Therefore, it is required that the scale error due to the waveplate rotator is on the order of 1\,\% or lower.

The tolerance of the error in the polarization-angle measurement is 2$^\circ$ for CLASP \citep{ishikawa2014b}.  
For the Stokes parameters with $Q/I = 1$ and $U/I = 0$, the polarization-angle deviation of 2$^\circ$ causes 
deviations of the Stokes parameters $\delta Q$ and $\delta U$ as $\delta Q/I = 1 - \cos (2\times 2 ^\circ ) \approx 0.002$ and $\delta U/I = \sin (2\times 2 ^\circ ) \approx 0.07$, respectively.
Therefore if $Q/I = 1$ and $U/I = 0$, the crosstalk can be written as $\delta U/(Q-\delta Q)$, and its tolerance is $<$7\,\%.  
Because the circular polarization is expected to be negligible in the case of the solar Lyman-$\alpha$ observation \citep{ishikawa2014b}, we consider only the crosstalk between $Q$ and $U$ here.

We developed a polarization modulation unit (PMU), \textit{i.e.} 
a system to rotate the half-waveplate \citep{ishikawa2013} continuously, comprising
a rotating motor (PMU-ROT) and its driver electronics (PMU-DRV).  
The PMU-ROT used for CLASP is the second-generation model developed for SOLAR-C. 
Its control circuit in the PMU-DRV is newly developed for the precise rotation.  
The PMU is designed to be robust for not only a rocket experiment but also space applications.
The CLASP experiment is an opportunity to demonstrate that the motor configuration and control technique can 
be used in space for SOLAR-C.
The long-term performance is an important factor for spacecraft applications, and it is tested separately \citep{shimizu2014}.  
In this article, we describe the specifications and control technique for the PMU and report the experimental results of the effect on the polarization measurement.

\section{Design of PMU} 
The PMU-ROT is a three-phase brushless direct-current motor and the PMU-DRV is designed to rotate the waveplate with a period of 4.8~seconds for CLASP \citep{kano2012}.    
The PMU maintains the rotation rate of 4.8~seconds per rotation by measuring the rotation angle using an optical encoder
and the time by counting clock-signal pulses from the central processor unit (CPU).  
The feedback gain and other parameters are optimized according to the rotation speed and moment of inertia of the flight configuration.  
Low-pass filters are used in the feedback loop to reduce the influence
of momentary disturbances.  

\begin{figure}[htbp]
 \begin{center}
 \includegraphics[width=80mm]{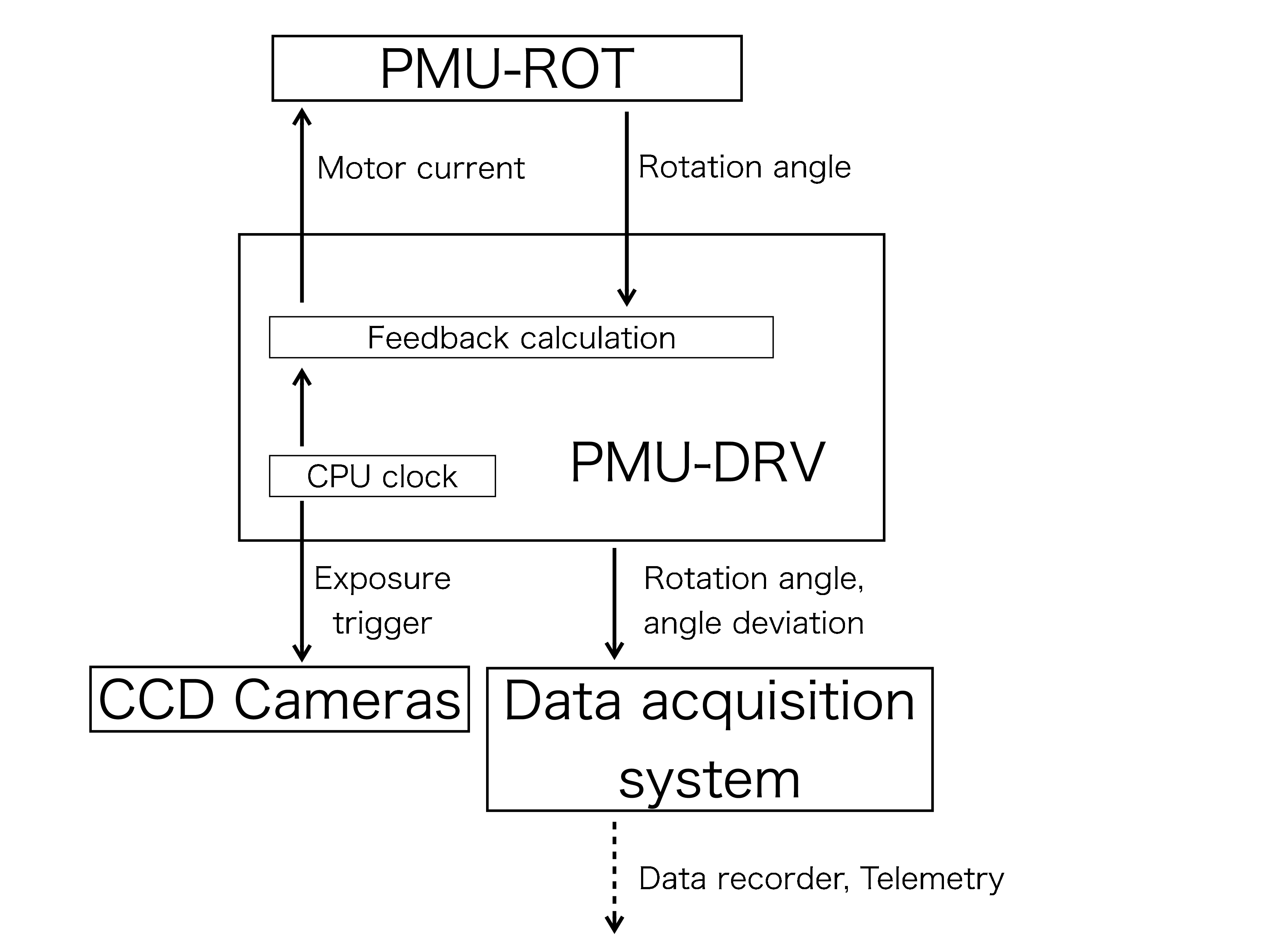}
 \end{center}
 \caption{Signal diagram of the PMU.  The PMU-DRV controls the PMU-ROT using the encoder signal, and 
 the PMU-DRV send signals to trigger the camera exposure and monitor the health and status of the rotation control.  
}
 \label{fig:signaldiagram}
\end{figure}

The timing of the exposure of the images should be synchronized with the rotation angle of the waveplate.  
If the exposure signal is generated by the angle measured using the encoder, 
the exposure duration may fluctuate because of the non-uniform rotation.  
Therefore, the PMU regularly calculates the target angle to maintain the uniform rotation based on the CPU clock in the PMU-DRV, and also 
generates an exposure synchronization signal regularly every 0.3~seconds by the same CPU clock, independently from the measured angle.  
Frame-transfer type charge coupled device (CCD) cameras receive the exposure synchronization signal.  
They then finish the ongoing exposures, transfer the frames, and start the next exposures.  
The PMU also outputs the measured rotation angle and angle deviation from the target angle to check the non-uniformity in the rotation.  
Those values are downlinked to the ground station by the telemetry during the flight.  
The signal diagram is shown in Figure~\ref{fig:signaldiagram}.

Materials with a low outgassing rate are selected for the PMU \citep{imada2011}. 
This is especially important for vacuum-ultraviolet observations.  
For CLASP, we baked out the PMU-ROT and confirmed using a thermoelectric quartz-crystal microbalance
that it exhibited no significant continued outgassing.

\section{Measurements of Rotation Uniformity and Influence on Polarization Measurement} 

We performed polarization modulation measurements to estimate the scale error and crosstalk for the flight instruments PMU-ROT and PMU-DRV.  
The uniform rotation of the PMU can be disturbed by momentary torque disturbances in the lubricant, the machining accuracy of the optical encoder, and the PMU-ROT structure itself.
The PMU-ROT has 11 magnets \citep{shimizu2014} which may cause periodic deviation of the rotation.  
Therefore, it is important to evaluate the scale error and crosstalk due to the PMU to confirm its performance.
 
\begin{figure}[htbp]
 \begin{center}
  \includegraphics[width=55mm]{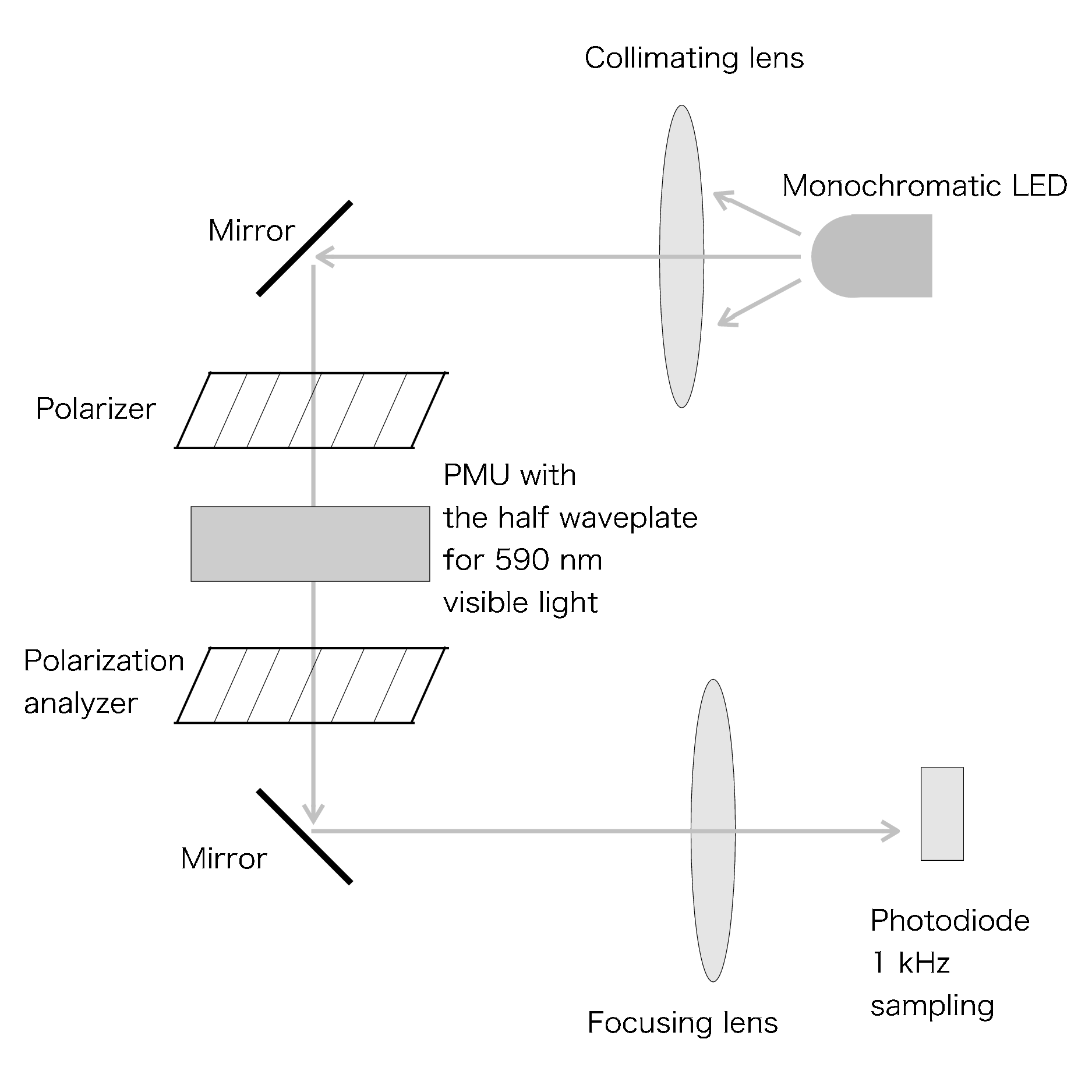}
  \includegraphics[width=65mm]{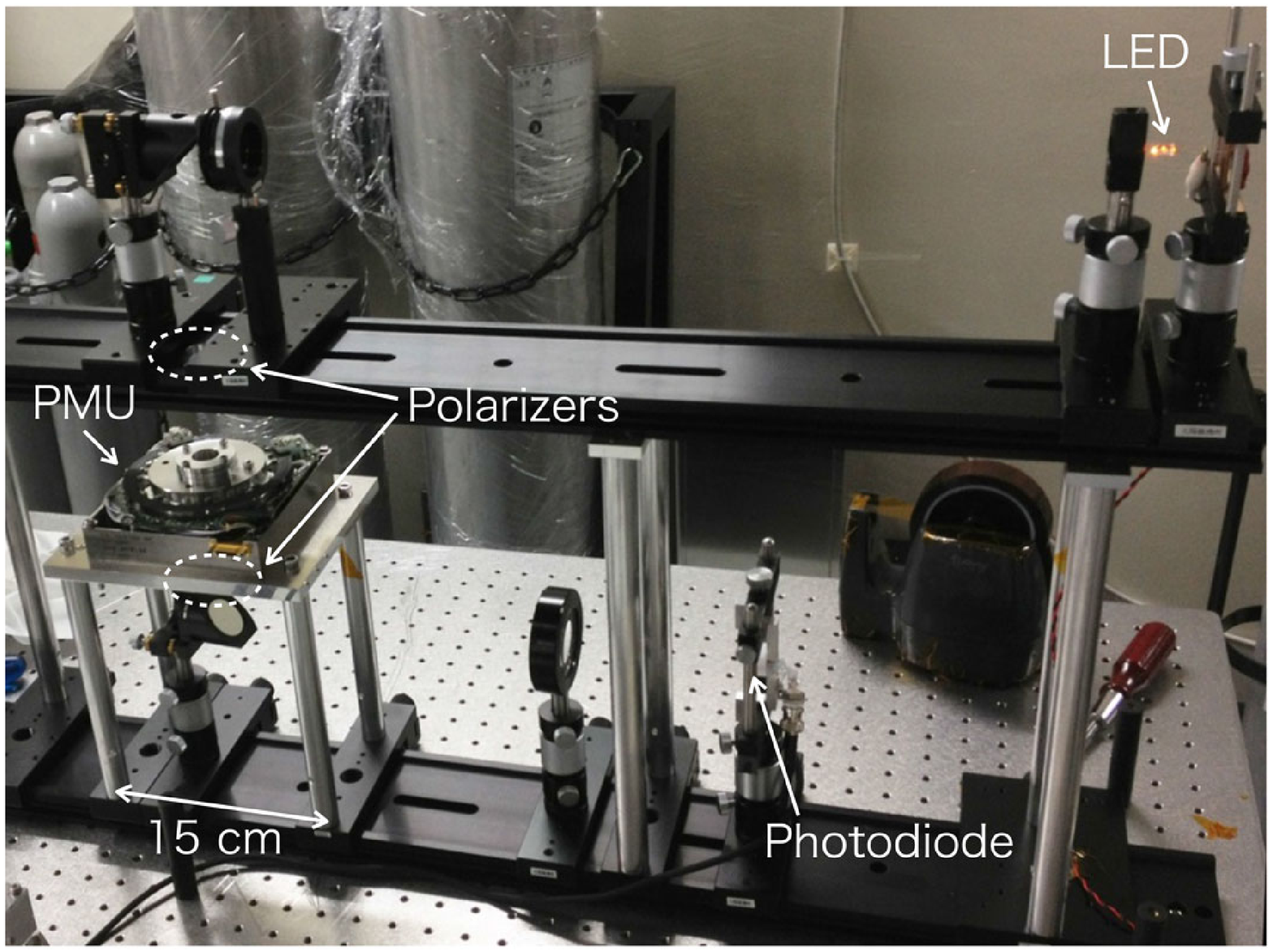}
 \end{center}
 \caption{(Left) Optical system diagram for the rotation-uniformity measurement.  
(Right) Photo of the experimental setup for the rotation-uniformity test.}
 \label{fig:rottest}
\end{figure}

The optical system of the experiment is shown in Figure~\ref{fig:rottest}.
A light-emitting diode (LED, EK Japan LK-5YE) emitting monochromatic light at a wavelength of 590~nm, with a wavelength width less than 10~nm, is used as a light source.
The variations of the luminosity are measured to be less than 0.1\,\% during the time scale of $<$1~minute.  
The LED light is collimated by a collimating lens and linearly polarized by a stack of two polarizers, and it is used as the incident beam for the rotating waveplate. 
The collimated linearly polarized light illuminates the rotating half-waveplate mounted on the PMU, 
which changes the direction of the linear polarization continuously.  
The waveplate used in this experiment is not for the Lyman-$\alpha$ line; rather, it is a commercial waveplate for visible light with a wavelength of 590~nm (Sigma Corporation WPQ-5900-2M).  
After passing the waveplate, the light passes through a stack of two polarization analyzers -- which are identical to the polarizers located upstream of the waveplate --
and is focused on a photodiode.
The polarizers before the waveplate and the polarization analyzers after the waveplate are aligned in the same direction, vertical to the incident and reflected lights. 
The intensity modulation is measured by a silicon photodiode (Hamamatsu Photonics S1337) at a sampling frequency of 1~kHz.  
The wavelength width of the light source and the wavelength dependence of the retardation of the waveplate cause a constant offset in the intensity modulation.  
The monochromatic LED light source is selected to minimize this offset.  
The experimental setup is covered by a blackout curtain, and the influence of stray light can be ignored.  
Although the observation duration for the sounding rocket 
is limited to $\approx$five~minutes, to evaluate the scale error and crosstalk, 
we measure the polarization modulation for $\approx$one~hour, corresponding to 737 rotations. 

The intensity modulation $\left[ i(t) \right]$ measured using the LED is described by the following equation: 
\begin{equation}
i (t) =  A \left[  \cos\left( 4\theta (t)\right)  +1  \right] + b, 
\end{equation}
where $A$ is the constant amplitude, $\theta (t)$ is the rotation angle, $t$ is time, and $b$ is the constant offset caused by the wavelength width of the light source and 
wavelength dependence of the retardation of the waveplate.  
The ideal intensity intensity $\left[ i _0 (t) \right]$ with uniform waveplate rotation at the rate $\left[ \omega \right]$ is given as
\begin{equation}
i_0 (t) =  A  \left[ \cos\left\{ 4(\omega t + \theta _0 ) \right\}  +1  \right]  + b,  \label{eq:ideal}
\end{equation}
where $\theta _0$ is the initial angle and $\omega$ is the angular velocity.  
In our experiment, we confirmed that the background intensity $\left[ b \right]$ is sufficiently smaller than $A$, and 
its time dependence is negligible.  
Therefore, the residual $\left[ \delta (t) \right]$ caused by the rotation non-uniformity can be written as
\begin{equation}
\delta (t) \equiv \frac{i(t)-i_0 (t)}{2A} =  \frac{\left[ \cos\left\{ 4(\theta (t)+ \theta _0 ) \right\}  - \cos\left\{ 4(\omega t + \theta _0 ) \right\} \right]}{2}. \label{eq:residual}
\end{equation}
Assuming that the angular deviation $\delta \theta (t) = \theta (t) - (\omega t + \theta _0)$ is sufficiently small, the relation between $\delta (t)$ and $\delta \theta (t)$ can be described as
\begin{eqnarray}
\delta (t) 
  \approx   - 2 \delta \theta (t) \sin  \left\{ 4(\omega t +  \theta _0 ) \right\}. \label{eq:angledev}
\end{eqnarray}

\begin{figure}[htbp]
 \begin{center}
 \includegraphics[width=100mm]{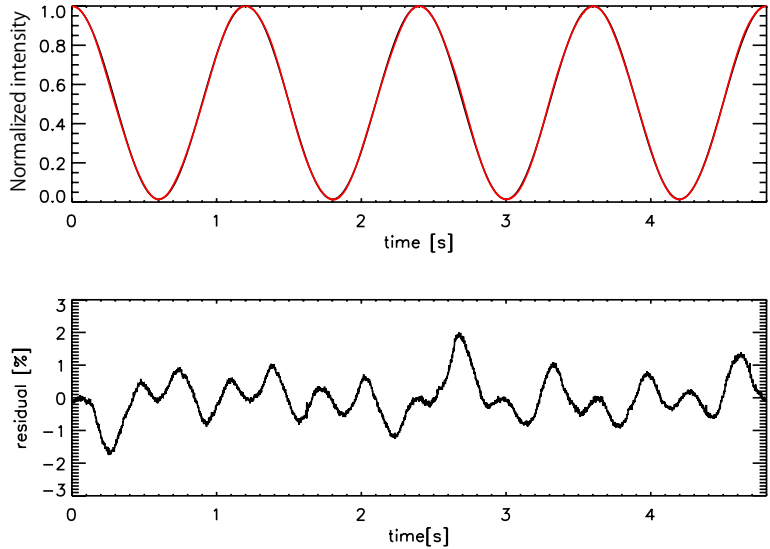}
 \end{center}
 \caption{Example of the measured polarization modulation with a single rotation.  Upper panel: measured polarization modulation (black) with the fitting curve (red).  
 Lower panel: residual between the measured modulation and fitting curve.  The residual is less than $\approx$2\,\% during the rotation.  
}
 \label{fig:modulation}
\end{figure}

Figure~\ref{fig:modulation} shows an example of the measured intensity modulation during a single rotation, according to the data measured continuously for one hour. 
As shown in the upper panel, 
the measured modulation and the fitted sinusoidal curve (black and red lines respectively) almost overlap each other.
The fitting is performed for each single rotation using the function given by Equation~(\ref{eq:ideal}), where the free parameters are $A$, $\theta _0$, and $b$.  
We use the fixed value of $\omega$ from the specification of the PMU.  
As shown in the lower panel, the residual $\left[ \delta (t) \right]$ between the measured and fitted curves 
is less than $\approx$2\,\% during the rotation, 
which corresponds roughly to an angular deviation of $\approx$1~$^{\circ}$, according to Equation~(\ref{eq:angledev}). 

 In the residual, we observe 15 peaks during one rotation and seven groups of them, 
each comprising two or three continuous peaks.  
 This pattern is observed at any time in the one-hour data, and is well explained by the structure of the PMU-ROT.  
Because the PMU-ROT has 11 magnets, as previously mentioned, it is expected that the angular deviation has a periodicity of 11 times per rotation.  
Accordingly, $\delta \theta (t) \propto \sin (11 \omega t + \theta '_0 )$, where $\theta ' _0$ is the initial phase of the magnets.  
Therefore, the product  of $\sin (11 \omega t + \theta '_0 )$ and $\sin \left\{ 4(\omega t + \theta _0 ) \right\}$ yields beats with 
periodicities of $15 \omega (= 11 \omega + 4 \omega)$ and $7 \omega (= 11\omega - 4\omega)$.

\begin{figure}[htbp]
 \begin{center}
 \includegraphics[width=56mm]{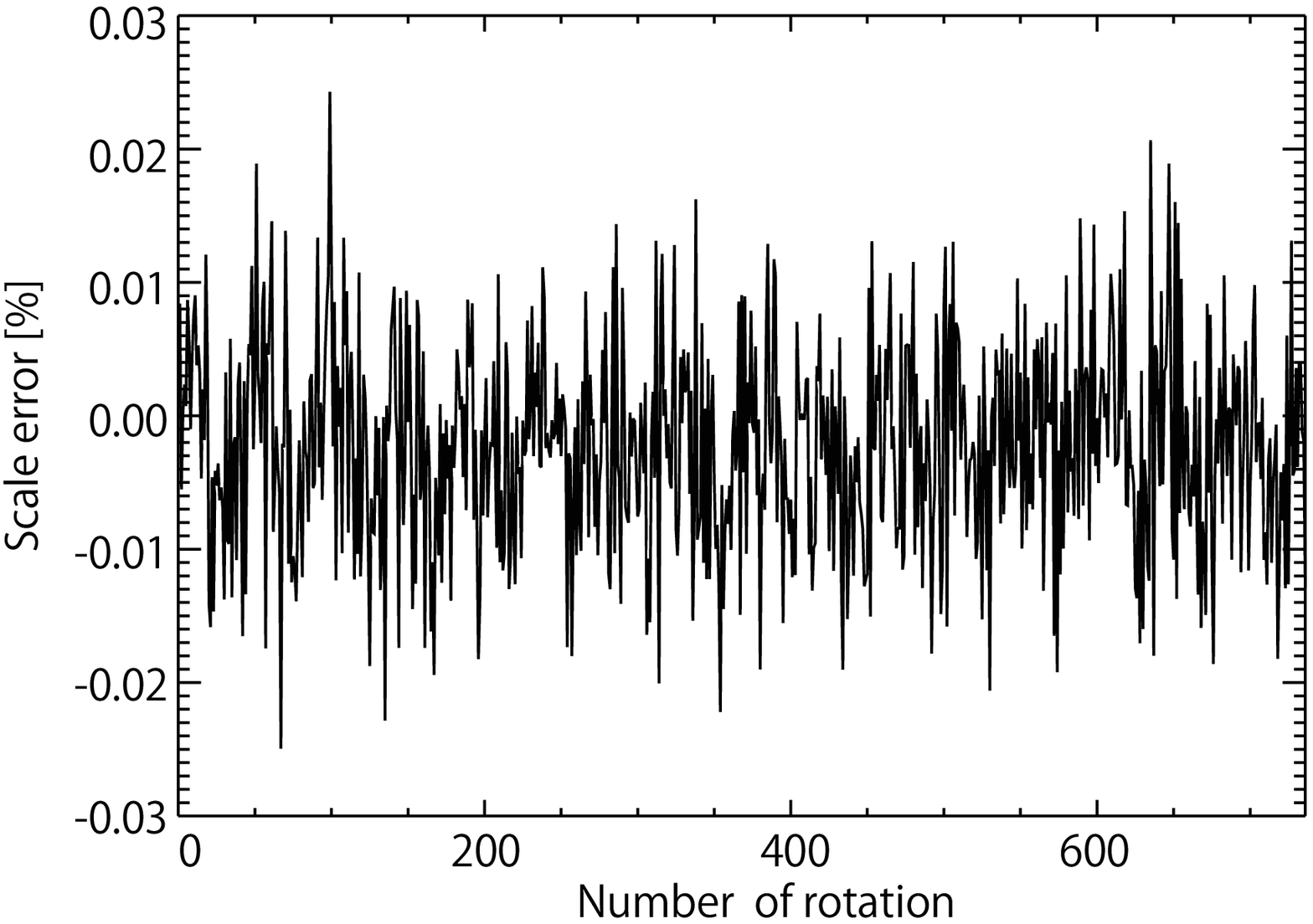}
 \includegraphics[width=60mm]{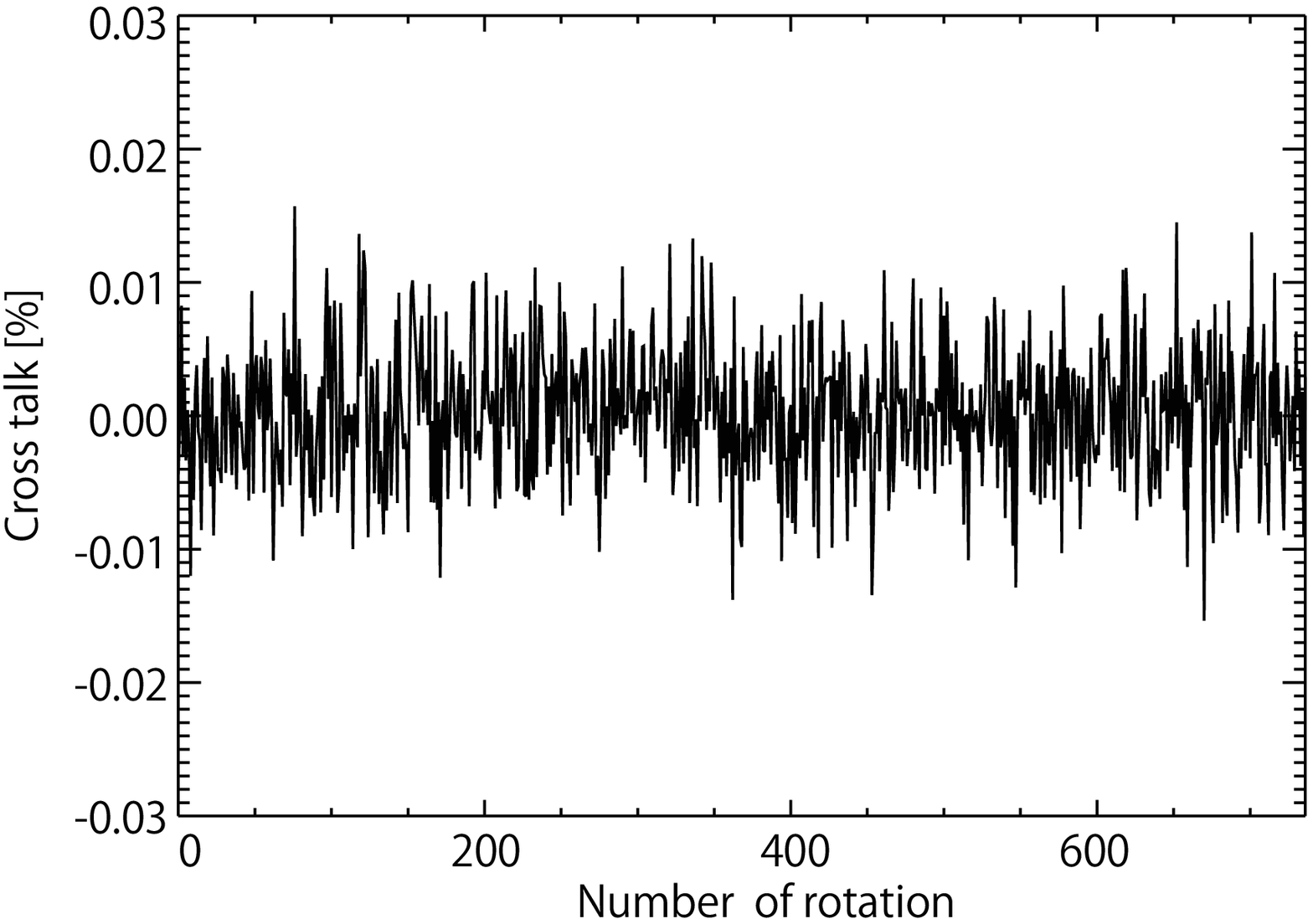}
 \end{center}
 \caption{Temporal variation of the scale error (left) and crosstalk (right) for the degree of polarization calculated by each rotation of the PMU.}
 \label{fig:scaleerror}
\end{figure}

To simulate the CCD camera operation during the flight, 
we integrated the intensity every 0.3~second (16 exposures per rotation) 
and then calculated the Stokes $Q/I$ and $U/I$ using the demodulation formula 
for the actual flight data (see Equation~(1) of \citealt{ishikawa2014b}).  
Similarly, we also calculated Stokes-$Q$ and -$U$ for the ideal rotation -- $Q_0$ and $U_0$ respectively -- using the fitted modulation curve. 
Because we extracted the data for each rotation from the local maximum in the intensity modulation, 
the data indicate that $Q _0 \gg U _0$ and $\theta _0 \approx 0$. 
Neither $U_0$ nor $\theta _0$ is exactly zero, as 
the sampling timings and peaks of the intensity modulation are not synchronized.  
In this situation, we can evaluate the scale error and crosstalk in the linear polarization as $(Q-Q_0)/Q_0$ and $(U-U_0)/Q_0$, respectively.  
The scale error and crosstalk for every single rotation in the one-hour operation are plotted in Figure~\ref{fig:scaleerror}.  
They were always less than 0.03\,\%.


The average and standard deviation of the scale error in the data for  the 737 rotations in the one-hour measurement are presented in Table~\ref{table}.  
\begin{table}
\caption{Averages and standard deviations of the scale error and crosstalk for the 1-h measurement.}
\label{table}
\begin{tabular}{ccc}     
  \hline                   
Component & Average [\%] & Scale error [\%] \\
  \hline
Scale error & 0.002 & 0.008\\
Cross talk & $<$ 0.001 & 0.005\\
  \hline
\end{tabular}
\end{table}
The average corresponds to the non-uniformity synchronized to the rotation period (as shown in the bottom panel of Figure~\ref{fig:modulation}), 
and the standard deviation mainly corresponds to the random fluctuation in the rotation, such as variations in the condition of the lubricant.
The 1$\sigma$ error of the scale error is less than $0.002\,\% + 0.008\,\% = 0.01\,\%$, and the 3$\sigma$ error is less than $0.002\,\% + 3 \times 0.008\,\% = 0.026\,\%$, which is small 
compared with the other error origins \citep{ishikawa2014b}.
Because the temporal variation of the scale error appears almost random around its average,
the scale error can be reduced by integrating more rotations to calculate the Stokes-$Q$.

The average and standard deviation of the crosstalk in the same data set are also shown in Table~\ref{table}.
As before, 
the 1$\sigma$ error of the scale error is less than $0.001\,\% + 0.005\,\% = 0.006\,\%$, and the 3$\sigma$ error is less than $0.001\,\% + 3 \times 0.005\,\% = 0.016\,\%$, 
which is also small. 
The crosstalk can also be reduced by integrating more rotations.  

   
\section{Conclusions} 
We developed a polarization modulation unit to rotate a waveplate continuously for precise spectropolarimetry.  
In measuring the linear polarization, 
we evaluated the scale error and crosstalk, finding that both are $\approx 0.01$\,\% or less (1$\sigma$ errors).  
This result shows that the scale error caused by the rotation non-uniformity has a far smaller influence than the other effects \citep{ishikawa2014b}.  
Both the scale error and the crosstalk satisfy the requirements for CLASP.  

\begin{acks}
We would like to thank H. Hara at the National Astronomical Obsevatory of Japan, S. Obara and K. Watanabe at JAXA, and S. Imada at Nagoya University for the development of the PMU-ROT.  
We also thank S. Hirata, M. Matsumoto and A. Urayama at the Mitsubishi Precision Co.,Ltd for the development of both of the PMU-ROT and PMU-DRV, and the Techno-Craft Co. for the development of the PMU-DRV.  
The PMU-ROT is developed with supports by the JAXA strategic research and development grant to the JAXA SOLAR-C working group.  
The CLASP sounding rocket experiment is founded by NASA, CNES, JAXA, and the Japan Society for the Promotion of Science (JSPS) through a Grant-in-Aid for Scientific Research (S) (Grant Number 25220703, PI: S. Tsuneta).
The development of CLASP in Japan was also supported by the basic research program of the Institute of Space and Astronomical Science (ISAS), internal research funding of the National Astronomical Observatory of Japan (NAOJ), and JSPS KAKENHI Grant Numbers 24340040, 23340052, and 24740134. The authors also acknowledge N. Okada at the Advanced Technology Center of NAOJ for his support in preparation of the experimental jigs.
\end{acks}

\section*{Disclosure of Potential Conflicts of Interest}
The authors declare that they have no conflicts of interest.



   

\bibliographystyle{spr-mp-sola}
\tracingmacros=2
\bibliography{clasp_pmu_s_ishikawa2015_bibliography}

\end{article} 

\end{document}